# Magnetic properties of magnetically textured Bi-2212 ceramics

Ph. Vanderbemden *[1], H. Bougrine [1], M. Ausloos [2] and R. Cloots [3].

[1] Montefiore Electricity Institute B28, [2] Physics Institute B5, [3] Chemistry Institute B6, University of Liège, S.U.P.R.A.S., Sart-Tilman, B-4000 Liège, Belgium.

## Abstract

This paper aims at reporting magnetic properties of bulk polycrystalline $Bi_2Sr_2Ca_{0.8}Dy_{0.2}Cu_2O_{8-y}$ samples textured under a magnetic field. The microstructure of these materials is highly anisotropic and exhibits particular features needed to be taken into account in order to interpret their magnetic and electrical properties. First the AC magnetic susceptibility $\chi = \chi' - j\chi''$ has been measured for several magnetic fields ($H // ab$ and $H // c$) and compared to the electrical resistivity data. The structure of the $\chi''$ peak is shown to be related to the chemical content distribution of the superconducting grains. Next, the magnetic flux profiles have been extracted from the magnetic measurements using the Campbell - Rollins procedure. The anisotropy of the flux profiles and their peculiar curvature behaviour for $H // c$ point out the role of both grain platelet structure and the presence of secondary phases. From these results, we conclude that the magnetic properties of such magnetically textured materials do not allow for a reliable extraction of the critical current density $J_c$ but essentially probe geometric effects. Such effects have to be taken into account for improving the manufacture of attractive high-$T_c$ materials.

*Keywords*: textured materials, magnetic properties, anisotropy

## 1. Introduction

The interpretation of magnetic measurements carried out on High-$T_c$ Superconductors often requires great care. The reason is that the magnetic response is caused by *intragranular* shielding currents, flowing in individual superconducting grains, as well as by *intergranular* currents, flowing across grain boundaries. In the case of textured materials, the anisotropy of the properties represents an additional parameter to be considered for understanding the measurement results. The purpose of the present paper is to bring out the relevant parameters affecting the magnetic properties of Bi-2212 textured materials and to present some appropriate experiments used for their determination.

---

*Corresponding author: tel: +32 4 366 2674; fax: +32 4 366 2877; e-mail : vanderbe@montefiore.ulg.ac.be

**2. Experimental techniques**

Several 2212 Bi-based textured samples have been prepared following an original technique developed in the SUPRAS group and described in ref. [1]. The method involves sintering and partial melting of 2212 Bi-based materials under an external magnetic field. The anisotropy of the grains paramagnetic susceptibility is enhanced by partially (20°%) substituting a rare-earth ion, like dysprosium, for calcium. The microstructure of such magnetically textured $Bi_2Sr_2Ca_{0.8}Dy_{0.2}Cu_2O_{8-y}$ samples indicates a significant alignment of the superconducting grains which exhibit a platelet structure along the ab planes. The platelet dimensions are about 5°x 200 x 200 m. An EDX analysis performed on the surface of polished specimens reveals that, despite an initial stoechiometric composition $Bi_2Sr_2Ca_{1-x}Dy_xCu_2O_8$ (x°=°0.2) before processing, the Dy ions do not substitute on the Ca sites in a uniform manner. The final samples result thus from some intergrowth of a Dy-free (x°=°0) and several Dy-rich (x°>°0.2) phases.

Magnetic measurements were performed in a home-made susceptometer based on a cryocooler [2]. The superconductor magnetic response was measured by a couple of orthogonal pick-up coils made of fine copper wire (50° m diameter) and wound close to the sample surface. Each pick-up coil signal was treated by an EG&G 5210 lock-in amplifier to determine the AC magnetic susceptibility and the magnetic flux profiles using the so-called Campbell technique [3].

**3. Results**

*3.1. Electrical resistivity and AC susceptibility*

Electrical resistivity vs. temperature curves are shown in Fig. 1 when the injection current is applied parallel (//) or perpendicular (⊥) to the ab-planes of the melt-textured sample. The phase-locking temperature, below which the resistivity vanishes, is approximately 82°K for both orientations. The AC magnetic susceptibility is plotted on the same graph for an external magnetic field of $_0H_{AC}$°=°1°G and f°=°1053°Hz applied either parallel/(/) or perpendicular (⊥) to the c-axis of the sample. The onset of magnetic susceptibility occurs *below* the phase-locking temperature for both orientations, indicating that the magnetic shielding is only caused by *intergranular* currents. Hence the properties discussed below are characteristic of the weak link network.

It is of interest to look at the out-of-phase component ($\chi''$) measured at different field amplitudes (Fig. 2). A first peak, occurring between 75°K and 83°K, is only visible for the H°//°c configuration. On lowering the temperature further, a second broad peak is observable for both field orientations. This suggests the presence of two kinds of weak links in these magnetically melt-textured materials. Some "*good* quality" links, hereafter called $L_G$, connect grains in the ab plane whereas "*bad* quality" links ($L_B$) connect the platelets through their large face. An overall schematic illustration of both weak links is proposed in the inset of Fig. 2. The $L_G$ links are only efficient when shielding currents flow in the ab planes, i.e. for the H°//°c configuration. The $L_B$ links allow the intergranular supercurrents to flow parallel to the c-axis. Therefore they contribute to the magnetic shielding for H°//°ab. However, because of percolation effects, shielding currents parallel to the ab planes (i.e. induced by a magnetic field parallel to the c-axis) can also flow through the $L_B$ links and improve the magnetic shielding along the c-direction. This reasonably explains why the $\chi''$ peak associated with the $L_B$ links is observable for both field orientations. In addition, the width of the $\chi''$ peak associated with the $L_B$ links is thought to be related to the distribution of critical temperatures of grains characterized by different Dy contents. An anisotropic weak link structure was also reported by Loughran and Goldfarb [4] in grain aligned YBCO ceramics. However, the anisotropy is much more emphasized in the present case, probably because of the platelet structure of the grains as well as the different chemical contents in our materials.

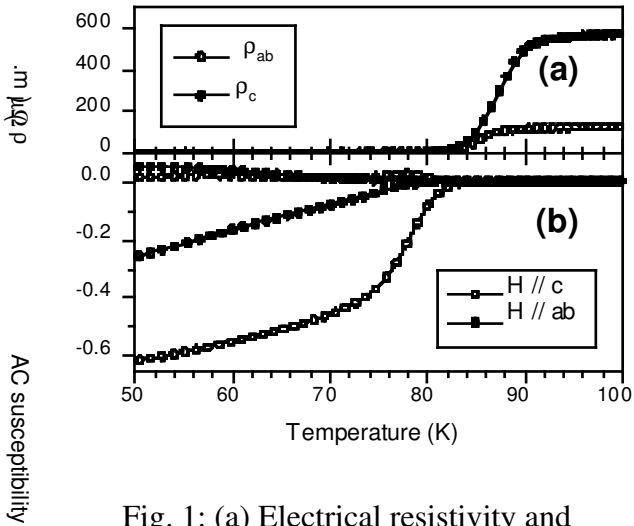
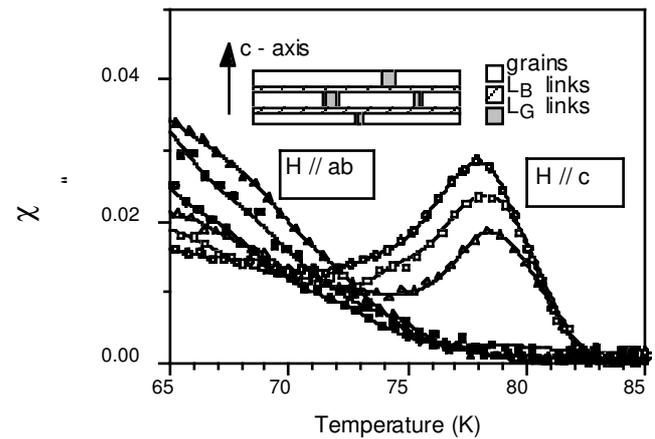

Fig. 1: (a) Electrical resistivity and (b) AC magnetic susceptibility as a function of temperature.

Fig. 2 : AC magnetic susceptibility out-of-phase component ($\chi''$) measured for several AC magnetic fields ($_0H = 1, 2$ and 5 G).

*3.2 Magnetic flux profiles*

The magnetic flux profiles, determined at a fixed temperature T°=°60°K, are shown in Fig. 3. For a magnetic field parallel to the c-axis of such textured samples, the flux profiles do not exhibit the so-called "kink feature" - i.e. a *positive* curvature - usually observed in sintered ceramics [5]. The flux profiles rather display a *negative* curvature, particularly visible for the curves corresponding to the highest fields. Since this effect is not present when H°//°ab, it is thought to be associated with strong demagnetizing effects arising from the flat geometry of the grains submitted to a transverse magnetic field (H°//°c). Indeed, according to SEM observations [1], the structure of the textured sample can be considered as an infinite array of thin superconducting strips. In such a geometry, according to analytical calculations worked out by Mawatari [6], the corresponding induction profiles B(x) have been shown to be non-linear and *formally* similar to those calculated for an infinite slab exhibiting a strong $J_c(B)$ dependence ($J_c^° \sim °B^{\pm}$). Qualitatively speaking, the flux profiles are therefore expected to follow a behaviour similar to those measured for non-textured ceramics in the absence of DC field, i.e. a *negative* curvature. Therefore it can be concluded that the behaviour observed in Fig. 3 has it signature in the platelet structure of the superconducting grains.

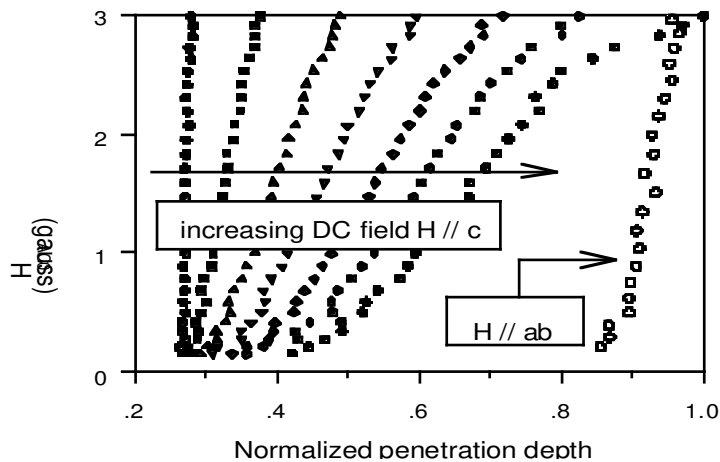

Fig. 3 : Flux profiles measured for several magnetic fields (0 to 90˚G). The temperature is fixed at 60˚K.

## 4. Conclusion

From the above considerations, it turns out that magnetic measurements on such magnetically textured materials essentially probe the *geometric* effects. Thus they do not allow for a reliable critical current determination. However the set of measurements presented here has brought out (i) the existence of different coupling strengths between grains, and (ii) the strong influence of the platelet structure of the superconducting grains on the magnetic properties.

## Acknowledgments

Ph. V. is grateful to the F.N.R.S. for the provision of a research grant. We also would like to acknowledge Prof. H.W. Vanderschueren for allowing us to use the M.I.EL. laboratory facilities. This work was part of an U.Lg. *Action de Recherches Concert es* (ARC 94-99 / 174) grant from the Science and Research Ministry of the *Communaut Fran aise de Belgique*.